\documentclass[apsrev4,pra,superscriptaddress,notitlepage,nofootinbib]{revtex4-1} 

\usepackage{amsmath,amsfonts,amssymb}
\usepackage{graphicx}
\usepackage[colorlinks=true, allcolors=blue]{hyperref}
\usepackage[version=4]{mhchem}
\begin{document}
\title{Photosensitive materials for neutron optics}
\author{Martin Fally}\thanks{Send correspondence to M.F.\\ E-mail: martin.fally@univie.ac.at}
\author{J\"urgen Klepp}
\affiliation{Faculty of Physics, University of Vienna, Austria}
\author{Christian Pruner}
\affiliation{Department of Materials Science and Physics, University of Salzburg, Salzburg, Austria}
\author{Elhoucine Hadden}
\affiliation{Faculty of Physics, University of Vienna, Austria}
\affiliation{Institut Laue-Langevin, Grenoble, France}
\affiliation{Vienna Doctoral School in Physics, University of Vienna, Austria}
\author{Andrea Bianco}
\affiliation{INAF-Osservatorio Astronomico di Brera, Merate, Italy}
\author{Joachim Kohlbrecher}
\affiliation{Paul Scherrer Institute, Villigen, Switzerland}
\author{Hanno Filter}
\author{Tobias Jenke}
\affiliation{Institut Laue-Langevin, Grenoble, France}
\author{Yasuo Tomita}
\affiliation{Department of Engineering Science, University of Electro-Communications,Tokyo, Japan}

\pagestyle{empty} 
\setcounter{page}{1} 
\newcommand{\bcr}{\Delta[b_c\rho]}
\newcommand{\bay}{Bayfol\textregistered HX\,}
\graphicspath{{/home/fallym4/MyConferences/24/PE24/LaTeX/fig/}}
\hypersetup{pdfauthor={Fally, Klepp, Pruner, Hadden, Bianco, Kohlbrecher, Filter, Jenke, Tomita},pdftitle={Photosensitive materials for neutron optics},pdfborder={0 0 0},urlcolor=blue}

\begin{abstract}
Photosensitive materials with ever-improving properties are of great importance for optical and photonics applications. Additionally, they are extremely useful for designing components for neutron optical devices. We provide an overview on materials that have been tested and successfully used to control beams of cold and very cold neutrons based on diffractive elements. Artificial gratings are generated and optimized for the specific application in mind. We discuss the needs of the neutron optics community and highlight the progress obtained during the last decade. Materials that have been employed so far along with their properties are summarized, outlining the most promising candidates for the construction of an interferometer for very cold neutrons.
\end{abstract}

\maketitle

\section{INTRODUCTION}
\label{sec:intro}  
Neutron interferometry using thermal neutrons at wavelength $\lambda=2$\AA\, was successfully demonstrated in 1974 by \textit{H. Rauch} and coworkers \cite{Rauch-pla74}. This led to a boost in neutron optics as it paved the way to tackle important fundamental physical questions by experiments. Among them the experimental proof of the $4\pi-$symmetry of the wave function for fermions \cite{Rauch-pla75,Werner-prl75,Rauch-81} must be highlighted as well as a number of measurements related to gravitational, topological and further quantum phase shifts \cite{Colella-prl75,Werner-prl79,Staudenmann-pra80,Summhammer-pra83,Greenberger-rmp83,Cimmino-prl89,Allman-prl92,Werner-cqg94,Werner-jpsj96,Allman-pra97}; for a comprehensive review see \cite{Rauch-15}. Recent experiments using an interferometer probe e.g. the quantum contextuality \cite{Bartosik-prl09}, the Heisenberg uncertainty principle \cite{Erhart-np12} or employ weak measurements \cite{Lemmel-prr22}.

All these phenomena rely on the \textit{phase} of a wave function, a quantity which can be accessed by an interferometer. Such instruments consist of three perfectly aligned diffraction elements which act as beamsplitters and mirrors. For thermal neutrons three slabs from a perfect silicon single crystal are used to Bragg diffract the interfering beams (triple-Laue arrangement of crystal slabs, see e.g. Ref.\,\citenum{Zawisky-nima02}). For neutrons with longer wavelengths, i.e. cold or very cold neutrons with $5\text{\AA}<\lambda<100\text{\AA}$, artificial gratings with grating spacings $\Lambda>\lambda$ are required. 

In this contribution we give a brief overview of materials that were tested for designing and producing artificial gratings. We elucidate the requirements to act as interferometer gratings, and discuss their properties and aptitude for manipulation of (very) cold neutrons. 

\section{Requirements for (very) cold neutron interferometer gratings}
The triple-Laue interferometer design employs the diffraction of neutrons from periodic structures to coherently split and reunite neutrons travelling along different paths (see a sketch, e.g. in Ref.\,\citenum{Lemmel-prb07}). The potential-dependent phase shift is given by the line integrals along the classical trajectories \cite{Rauch-15}. Thus its sensitivity in most cases depends on the enclosed area between these paths which is determined by the length of the interferometer and the diffraction angles. In an ideal case the first (G1) and third grating (G3) are 50:50 beamsplitters, the second grating (G2) serves as mirror. Deviations will lead to a reduced contrast of the interferometric signal. 

Therefore, the first requirement is a diffractive power for each individual grating, i.e. a diffraction efficiency $\eta_0=\eta_j=0.5$ (G1, G3) and $\eta_j=1$ in an ideal case,  where $j$ denotes the diffraction order. This is where various materials come into play. 

The second requirement is to maximize the diffraction. This forces us to make the grating spacings $\Lambda$ as short as possible.

A third desired property of the gratings is a low angular selectivity, while still preserving the diffractive power. This is a pragmatic necessity as an ultra-precise alignment of the gratings in the interferometer has to be achieved and very cold neutron sources provide a rather broad wavelength spectrum. 
\section{Neutron interferometers based on artificial gratings}
A first neutron interferometer for very cold neutrons was successfully operated after more than a decade of diligent construction based on three ruled gratings with a grating spacing of $\Lambda=2\,\mu$m and a length of $L=1.02\,$m \cite{Gruber-pla89,Eder-nima89,Zouw-nima00}, diffraction angles thus were about 5 mrad. The gratings had a duty ratio of $1$ and were of rectangular shape with thicknesses ("height") of 590 nm (G1, G3) and 750 nm (G2), respectively. These gratings with efficiencies around $\eta\approx 0.4$ operate in the thin grating regime \cite{Gaylord-ao81}. In particular this means that the neutron beam is diffracted simultaneously to several diffraction orders which is detrimental in two aspects: (1) the parasitic beams, not contributing to the interferometric signal, add up to a considerable background if not shielded from the detector (2) given the low number of very cold neutrons losses to unwanted directions must be avoided.

Neutron interferometers for the cold wavelength regime ($\lambda\approx 1\,$nm) were built based on recording artificial gratings by light optical holography in neutron-photosensitive materials. The tedious alignment process was accomplished during the recording process once for ever \cite{Schellhorn-pb97,Pruner-nima06}. The chosen material was deuterated photosensitized polymethylmethacrylate. Grating thicknesses were 2.7 mm, therefore operating in the thick grating or Bragg-regime \cite{Gaylord-ao81}: only two orders propagate, provided that the Bragg condition is fulfilled and diffraction efficiencies are high:
\begin{equation}
\label{eq:nDE}
\eta_1=1-\eta_0=\sin^2\left(\frac{1}{2}d\lambda\bcr_1\right).
\end{equation}
Here, $d$ is the thickness of the grating and $\bcr_1$ the first order coherent scattering length density modulation, where $b_c\rho$ is the product of the coherent scattering length $b_c$ - characteristic for the scattering power of a particular isotope - and the number density $\rho$. The light-induced changes form a one-dimensional grating and can be expanded in a Fourier series 
$\bcr(x)=\sum\limits_{s=-\infty}^\infty \bcr_s\exp{(\imath s 2\pi x/\Lambda)}$.

Diffraction in the Bragg-regime solves both of the issues addressed above. Unfortunately, the angular selectivity, which is proportional to the inverse thickness, is high in this case. To overcome also this problem our aim is to reduce the thickness $d$ while increasing $\bcr_1$ at the same time. The latter is decisive and the central material property discussed in this article.

\section{MATERIALS: PHOTOPOLYMERS}
The idea to employ photosensitive materials together with light optical holography for preparing diffractive elements for cold neutrons dates back to 1990 \cite{Rupp-prl90}. It was shown that preparing an optically thick grating ($d\approx 2\,$mm) in polymethylmethacrylate (PMMA) allowed to diffract neutrons \cite{Matull-zpb90}. In this case it is obvious that the coherent scattering length density modulation $\bcr$ originates only from density changes, i.e., $\bcr=[b_c\Delta\rho]$. 
While PMMA works fine for light a major disadvantage is the strong incoherent scattering originating from the prevalent \ce{^1H} isotopes. This serious issue was solved by using its deuterated analogue, d-PMMA \cite{Matull-el91} which was later used for the interferometer setup, too \cite{Schellhorn-pb97}. While d-PMMA proved its aptitude to diffract neutrons, it is not too easy to prepare and exhibits a long temporal evolution after recording was stopped (postpolymerization) \cite{Havermeyer-prl98,Rupp-osa99a}. Therefore, other photopolymers were explored. We will briefly introduce two interesting candidates: the commercially available \bay\cite{Bruder-Polymers17} and a recently synthesized CAS based polymer \cite{Galli-jps21}.

As polymers generally contain lots of hydrogen, and deuteration is technically demanding, we take advantage of the fact that the thicknesses we are attempting are by 1-2 orders of magnitude lower ($d<100\,\mu$m) and so is the incoherent scattering. Deuteration thus remains a feature nice-to-have but is not mandatory.
\subsection{\bay}
\bay is a commercially available photopolymer which is particularly optimized for easy recording of holograms, both of transmission as well as of reflection type. Standard sheets of the photopolymer have a thickness of the photosensitive layer of $d_0=16\,\mu$m supported by a $50 -- 100\,\mu\text{m}$ polymeric substrate. Thus a stack with doubled photopolymer layer thickness $2d_0$ between two substrates can be assembled \cite{Lahijani-spie23} and is ready for grating recording. This is performed by employing a standard two-wave mixing setup. 
Two coherent laser beams are brought to interference at the sample position which leads to a cosinusoidal variation of the light intensity. Due to the nonlinear optical response of \bay the light interference pattern is transferred in a neutron refractive-index modulation which serves as a grating even allowing for higher harmonics \cite{Bruder-spie15}. This is a tentative asset for increasing the (Bragg) diffraction angle $\Theta_j=\arcsin(j\lambda/(2\Lambda))$ when aiming at gratings for a neutron interferometer.

\begin{figure} [ht]
   \begin{center}
   \begin{tabular}{cc} 
   \frame{\includegraphics[height=7cm,trim=0 0 400 0,clip]{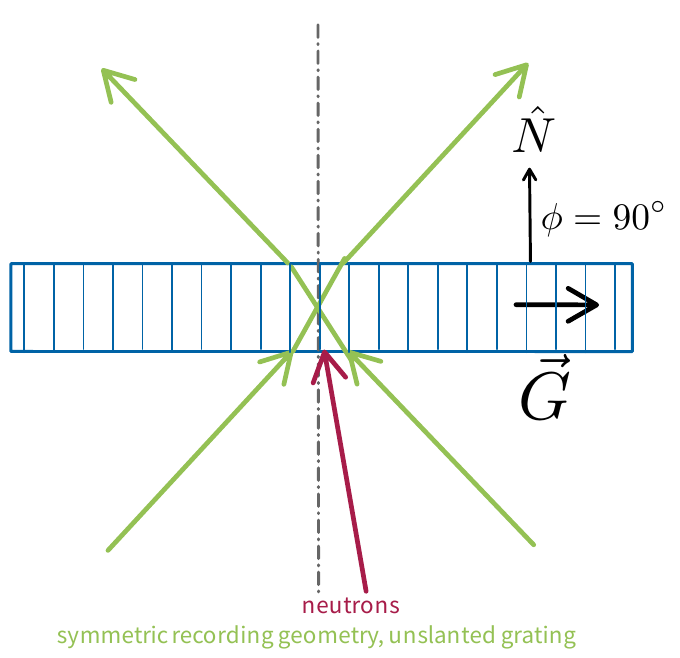}} & 
   \frame{\includegraphics[height=7cm,trim=470 0 0 0,clip]{geo_2_0-cropped}}
   \end{tabular}
   \end{center}
   \caption[IFsketch] 
   { \label{fig:geom} \textit{Left panel}: symmetric recording in transmission geometry (unslanted grating, $\phi=90^\circ$). \textit{Right panel}: aysmmetric recording in reflection geometry (slanted grating, $\phi\neq 0^\circ, 90^\circ$). Readout with neutrons is conducted in transmission for both cases.}
   \end{figure}

For the same reason and as the material is known to have an extremely high resolution we explored two different recording geometries (see \ref{fig:geom}):
\begin{enumerate}
	\item[(a)] the photopolymer was placed in a fully symmetric position with its sample surface normal along the bisectrix of the recording beams. Thus an unslanted holographic transmission grating for light and neutrons was obtained. The grating vector $\vec G$ then is perpendicular to the sample surface normal $\hat N$.
	\item[(b)] the photopolymer was placed in a slanted reflection geometry. This results in a slanted holographic reflection grating which then can be used as a slanted \textbf{transmission grating for neutrons}. The major advantage is a considerably decreased grating spacing $\Lambda=2\pi/|\vec G|$.
\end{enumerate}

Neutron diffraction experiments to evaluate the aptitude of \bay for neutron optical gratings were conducted at the beamline PF2/VCN at the high flux reactor of the Institut Laue-Langevin (ILL) in Grenoble. Data are made available by the ILL \cite{Klepp-ddoi20,Hadden-ddoi23}. A typical SANS-setup with a divergence of about 1 mrad and a broad wavelength distribution was used to examine the angular dependence of the diffraction efficiencies aka. rocking curve.

The results for the standard transmission grating of $\Lambda=491\,$nm for geometry (a) could demonstrate diffraction of neutrons with a coherent scattering length density modulation of only $\bcr_1=2.8\,\mu\text{m}^{-2}$ which is a rather low value as compared to other materials \cite{Lahijani-spie23}. This resulted in peak diffraction efficiencies for the $\pm$first orders of $\eta_{\pm 1}=4.5$\% at a mean wavelength $\overline{\lambda}=5.47\,$nm.

For geometry (b) we could demonstrate first order diffraction for neutrons from the slanted grating and validate the estimated grating spacing resulting from light optical reflection geometry to be $\Lambda\approx 177\,$nm. This short period is an important step for use in a VCN neutron interferometer with regard to enlarging the diffraction angle.
The diffraction efficiency, however, unfortunately was extremely low. Therefore, expected higher harmonics with even shorter spacings below 100\,nm could not be observed. Evaluation is still ongoing, more detailed results will be published elsewhere \cite{Hadden-PhD24}. 

Summarizing, \bay has the advantages of excellent optical performance, a strong nonlinearity of the recording process with distinguished higher harmonics and a high resolution. The disadvantag regarding their use in neutron optics is that the achieved coherent scattering length density modulation values are around $\bcr_1\approx 3\mu\text{m}^{-2}$ and below that of other materials.

\subsection{Cyclic allylic sulfide}
Only recently a new photopolymer system based on cyclic allylic sulfide (CAS) with addition of a thiol crosslinker was introduced as a new holographic material \cite{Galli-jps21}. The most important figure of merit, the refractive index modulation obtained after recording a grating, shows values of up to $\Delta n>0.03$ in the visible spectral range. As we have learnt in the previous section this is a prerequisite but not a garantuee for a large $\bcr$, the figure of merit for neutrons. 

A very important finding communicated in the publication mentioned above is, that a post thermal treatment improves the holographic grating further, viz. $\Delta n$ is even doubled. This increase is explained by a separation of the recording chemistry from the binder accompanied by a density increase. The latter gives strong evidence that diffraction of neutrons from such a grating could be prominent. 

First light optical diffraction experiments ($\lambda=543\,$nm, $\Lambda=838\,$nm, $d\approx 19\,\mu$m) were carried out to probe tentative higher harmonics which were not investigated in Ref.\,\citenum{Galli-jps21}. As discussed above, their existence would be a beneficial property whenever larger diffraction angles are required \cite{Havermeyer-prl98}.

\begin{figure} [ht]
   \begin{center}
   \begin{tabular}{p{\columnwidth/2}p{\columnwidth/2-5mm}} 
\frame{\includegraphics[height=6cm]{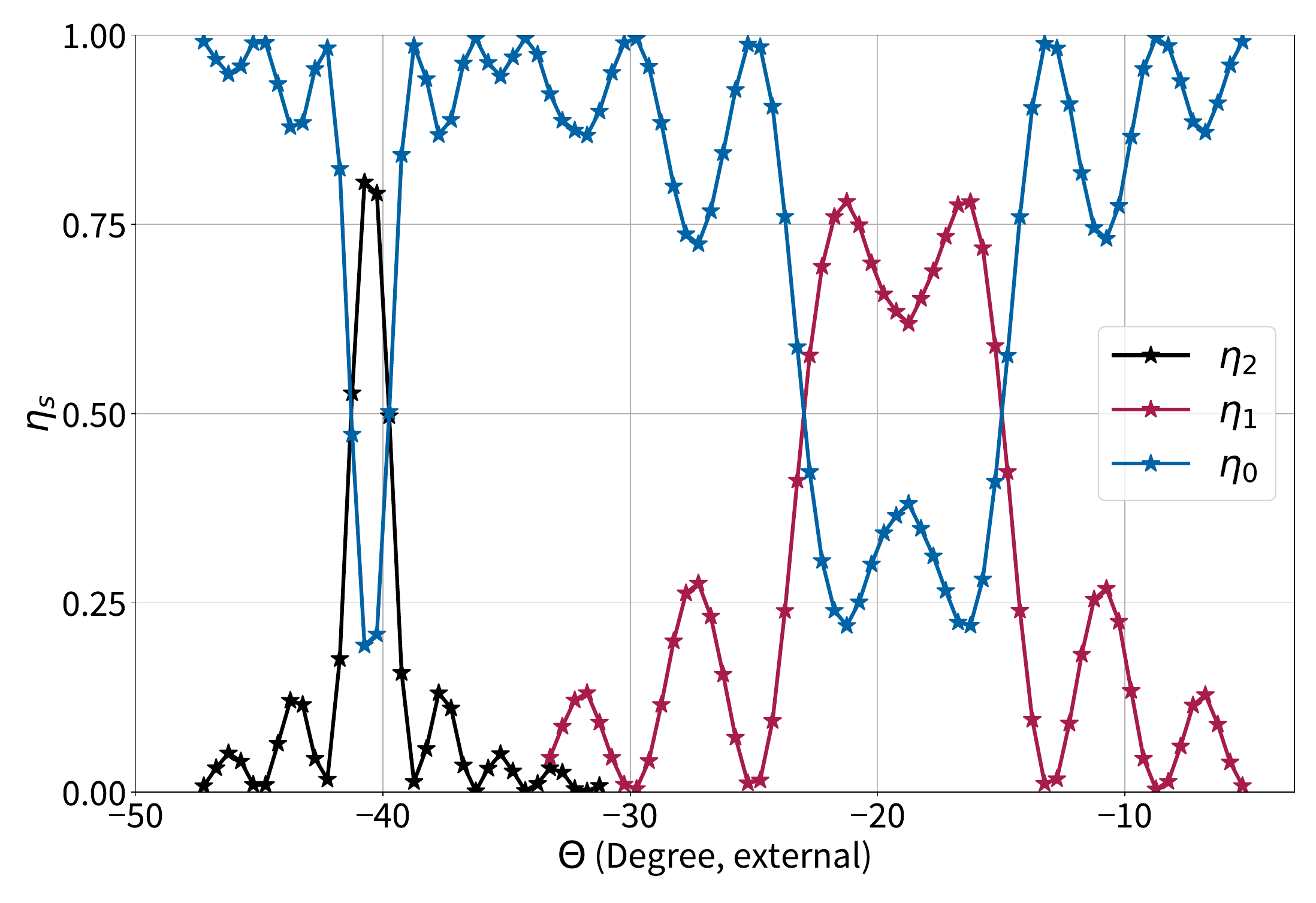}} &
\includegraphics[height=6cm]{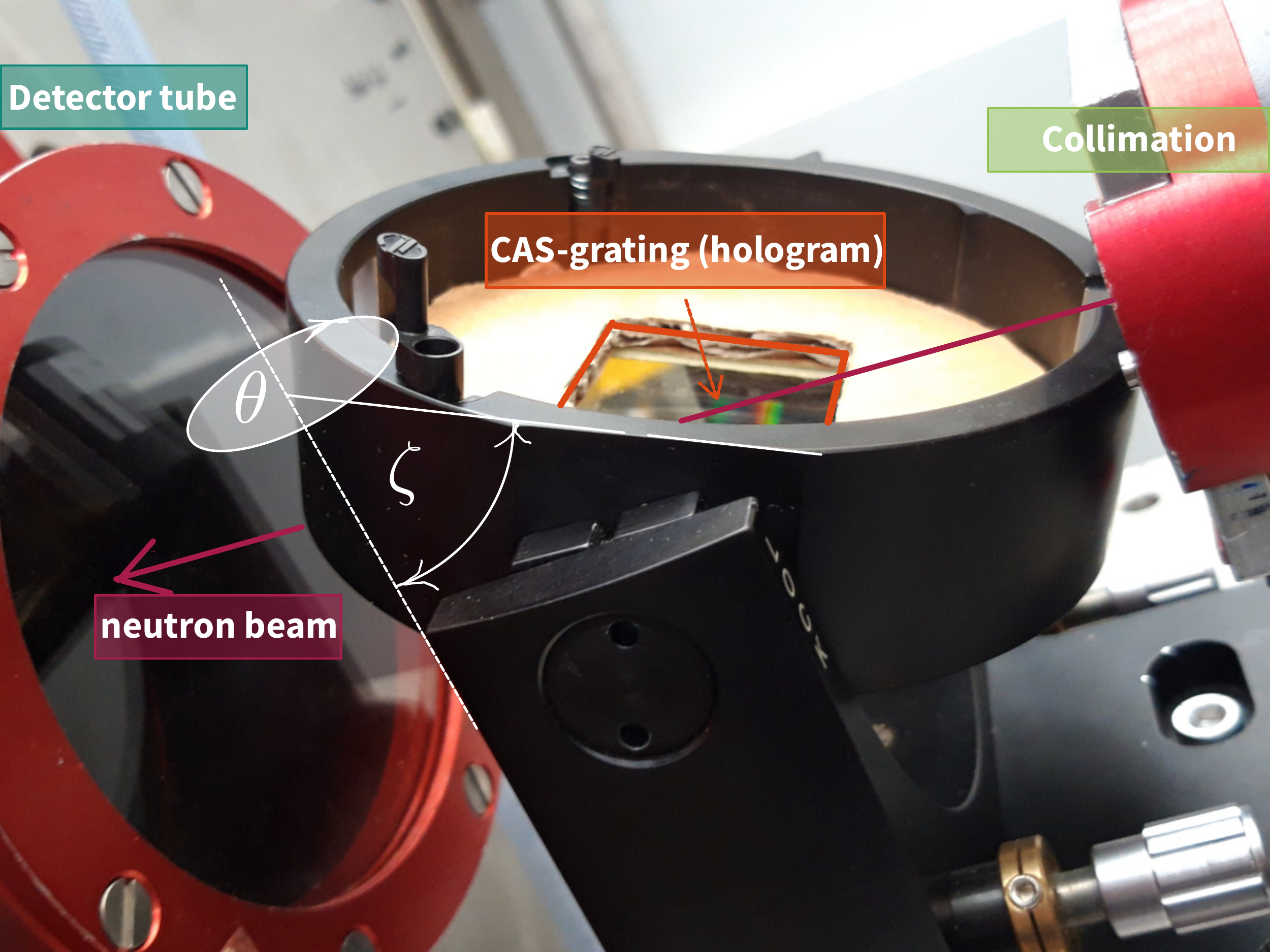}
   \end{tabular}
   \end{center}
   \caption[IFsketch] 
   { \label{fig:DE2} \textit{Left panel}: angular dependence of the light optical diffraction efficiency $\eta_ {0,1,2}(\Theta)$ for the zero, first, and second orders from a grating recorded in CAS. $\Theta$ are external angles.
   \textit{Right panel}: picture of the grating tilted at $\zeta\approx60^\circ$ at the neutron beamline (SANS, PSI). $\theta$ is the rotation angle around the axis perpendicular to $\vec G$.}
   \end{figure}

Indeed strong second order diffraction was observed with a corresponding refractive index modulation $\Delta n_2\approx 0.02$ evaluated by a rigorous coupled wave analysis \cite{Moharam-josaa95}. The angular dependence of the diffraction efficiency for light is shown in Fig.\,\ref{fig:DE2}.

Neutron optic diffraction experiments were carried out at the SANS beamline at the Paul-Scherrer Institut (PSI), Villingen. We chose a central wavelength $\lambda=2.62\,$ nm with $\Delta\lambda/\lambda\approx 0.1$ and a divergence of the beam of less than 2 mrad. To enhance the effective thickness $d$ of the grating and hence its diffraction efficiency we rotated the grating around an axis parallel to the grating vector by about a tilt-angle $\zeta=60^\circ$ which actually doubles the effective thickness $d=d_0/\cos(\zeta)=38\,\mu$m; see the picture of the setup in Fig.\,\ref{fig:DE2}. Subsequently the angular dependence of the diffraction efficiency around an axis $\theta$ perpendicular to the grating vector was probed.

The grating actually showed diffraction to the first and also second orders. The Bragg angles are small, $\theta_1\approx\lambda/(2\Lambda)=1.56\,$mrad and $\theta_2=3.13\,$mrad, whereas the width of the rocking curve is - as desired - broad. Its width can be roughly estimated by $\delta\theta=\Lambda/d=22\,$mrad. Thus all diffraction orders are present at once at angles of maximum diffraction efficiency near normal incidence (multiwave coupling). The measured intensities lead to diffraction efficiencies of around $\eta_{1}\approx 8$\% and $\eta_{2}< 0.2$\%. A thorough anaylsis of the data will be published elsewhere.

To get a glimpse on the relevant parameter $\bcr_1=b_c\Delta\rho_1$ for the first diffraction order we neglect the second orders and due to the low efficiency of the first orders we dare to employ the standard two-wave coupling theory \cite{Kogelnik-atj69}. Employing Eq.\,(\ref{eq:nDE}) this results in $\bcr_1\approx 2\sqrt{\eta_1}/(d\lambda)=5.9\,\mu\text{m}^{-2}$ which is among the largest values for materials investigated so far; e.g. would this grating at the typical VCN-wavelength of $\lambda=5\,$nm reach a diffraction efficiency of 30\%, a slightly thicker one with still low angular selectivity would then act as 50:50 beamsplitter as desired. This makes the material one of the top candidates for interferometer gratings, provided that (1) the grating spacing can be reduced without considerable loss of $\bcr$, and (2) a doubled geometrical thickness can be reached. 

As for the thickness, it is a question of changing the dye concentration and optimizing the deposition conditions to preserve the optical quality. Concerning the spacing, it was reported that a decrease of $\Delta n$ for light optics upon a decreasing grating spacing takes place \cite{Suzuki-ao04,Kowalski-jpsb16}. Therefore, a compromise between decreasing the grating spacing and retaining $\Delta n$ high must be made.
\section{MATERIALS: COMPOSITE MATERIALS}
The very beginning of applying composite materials was in 2005 when we tested holographic polymer dispersed liquid crystals (LCs), a hot top in nonlinear light optics at that time, for neutrons. The expectations were low as these materials had a thickness of a few tens of microns in compared to the several millimeters in d-PMMA and the diffraction efficiency depends quadratically on $d$ for low values according to Eq.\,\ref{eq:nDE}. Yet the light-induced phase separation between the LC and the polymer regions also would give rise to a density modulation which is also due to a separation of different species that have their individual $b_c\rho$. This was experimentally demonstrated for gratings, not particularly prepared for neutron diffraction, which had a grating spacing of $\Lambda=1.2\,\mu$m and a thickness of $d=30\,\mu$m at a cold neutron beamline ($\lambda=1.16\,$nm) \cite{Fally-prl06,DrevensekOlenik-spie07}. The obtained coherent scattering density modulation for the first order was impressive: $\bcr_1\approx 10\,\mu\text{m}^{-2}$. This type of material was a great step forward, however, we did not follow it for three reasons: (1) the thickness for recording was limited as the LCs scatter light strongly and destroy the interference pattern after a few ten microns propagation; (2) the LC component disappeared from the market; (3) an inevitable reduction of $\Lambda$ in the submicron range turned out to go along with a strong decrease of the phase separation \cite{Fally-joa09}. 
\subsection{Nanoparticle polymer composites}
One of the major innovations in holographic grating fabrication on the material side during the last decade was the development of nanoparticle polymer composites (NPC) with numerous different types of nanoparticles, polymeric components, stabilizers, photoinitiators and more \cite{Suzuki-apl02,Suzuki-jjap03,Suzuki-ao04,Smirnova-apb05,Chambers-n07,Sakhno-n07,Goldenberg-cm08,Gyergyek-csa08,Matras-mclc08,Tomita-ol08,Makovec-csa09,Ninjbadgar-afm09,Sakhno-joa09,Smirnova-n09,Gyergyek-csa10,Fujii-ol14}; cf. to Ref.\,\citenum{Tomita-jmo16} for a review. Some of these materials were tested, characterized and applied also for neutron optic purposes \cite{Fally-prl10,Klepp-nima11,Klepp-pra11,Klepp-apl12,Klepp-apl12a,Klepp-m12,Guo-ol14} and only recently nanodiamond polymer-composites were even designed for use in neutron optics \cite{Fally-spie20,Tomita-pra20,Tomita-spie20,Hadden-spie22,Hadden-apl24,Hadden-PhD24}.

Some of the most important advantages of using NPCs are:
\begin{itemize}
	\item their versatility concerning the species of nanoparticles having their individual \textbf{interaction with neutrons}, also allowing for spin dependent magnetic scattering in addition to the usually prevalent nuclear scattering
	\item the possibility to increase or decrease the interaction with the neutron by modifing their concentration
	\item their positive effects on the thermal \cite{Tomita-ol08} as well as the mechanical stability \cite{Suzuki-oex06,Hata-ol10}
\end{itemize}

It is actually possible to tune the $\bcr$ by changing the nanoparticle concentration and/or the species. The interaction with the polymeric matrix usually is very similar for the variety of components used. Incorporating nanoparticles allows to either enhance or to reduce the grating contrast. It is a peculiar property that most of the isotopes exhibit positive coherent scattering lengths and thus $b_c\rho$ while a few of them do have negative ones, e.g., \ce{^1H}, \ce{^7Li}, \ce{^48Ti}. Most of the materials are used "as-is", i.e., the polymers contain hydrogen in a natural abundance (except for seldomly used and expensive deuterated PMMA) of isotopes as well as the nanoparticles do. This implies that incoherent scattering ($4\pi$-scattering without information on the structure) occurs and adds to the unwanted background. It originates from either isotope incoherence (multiple isotopes with their individual $b_c$) or spin incoherence (interaction of the neutron spin and the nuclear spin) or both. Obviously, this deteriorates measurements and should be avoided by clever choice of the material. The aim is to maximize the contrast between $b_c\rho$ in the dark and the illuminated regions while minimizing potential incoherent scattering. Nanodiamonds fulfill this requirement: (1) they have a relatively high $b_c\rho$ which differs a lot from that of a polymer matrix, and (2) have low incoherent scattering and negligible absorption. The latter is of utmost importance as the absorption increases exponentially with wavelength and we employ the long VCN wavlengths. The effort of preparing such gratings pays off and let us reach the goal to provide gratings suitable for VCN-interferometers, i.e.: high diffraction efficiencies ($\approx 50$\% or $\approx 100$\%) at low angular selectivity (thin, $d<50\,\mu$m) providing comparably large diffraction angles (small grating spacing $\Lambda\leq 500\,$nm) with no extra background. One drawback of the nanodiamonds is related to their light optical properties: they are black. Therefore, the thickness of the sample must not exceed a few tens of microns because of heavy absorption losses and scattering, similar to the previously discussed holographic polymer dispersed LCs.

The experimental details on the realization of this type of nanodiamond polymer composite gratings are provided in Ref.\,\citenum{Tomita-pra20} with improved properties reported only recently \cite{Hadden-apl24}. Experiments using a SANS setup at the PF2/VCN beamline showed diffraction efficiencies of $\eta_1\approx 70\%$ and $\eta_2\approx 30\%$ at their corresponding Bragg angles. The coherent scattering length densities for the first and the second orders extracted from these data are remarkable even if an attenuation of the grating along the depth is observed: $\bcr_1=11.7\,\mu\text{m}^{-2}$ and $\bcr_2=5.2\,\mu\text{m}^{-2}$, respectively. Currently, nanodiamond-polymer composites are the gold standard for preparing holographic gratings as neutron diffraction optical elements for VCN. Further optimization is under way.

Recent tests also involved NPC gratings with Hyperbranched Polymers (HPBs) as organic nanoparticles. The preliminary results are still under evaluation \cite{Hadden-PhD24} and will not be covered in this work.

\subsection{Ionic liquid - photopolymer composites}
Another attempt we took was to prepare ionic liquid - photopolymer composites according to the formulation given in Refs.\,\citenum{Lin-apl08,Lin-om11}. The quality of the recorded gratings turned out to be inhomogeneous across an area of a square millimeter which is non-desirable, makes the neutron experiments a tedious task. The rather modest quality might originate from different non-ideal factors, among them the recording using a UV-laser source which promotes unwanted scattering during recording. Gratings with two largely different grating spacings $\Lambda= 480\,$nm or $\Lambda\approx 6\,\mu\text{m}$ for different recording conditions (exposure time) and thicknesses $d=10\ldots 100\,\mu$m were comprehensively studied \cite{Ellabban-m17}. 

The light optical measurements revealed a substantial attenuation of the complex refractive-index modulation along the grating depth, i.e., mixed, attenuated gratings \cite{Ellabban-spie17}.

Diffraction of neutrons from gratings with $\Lambda= 0.48\,\mu$m was performed at the PF2/VCN beamline using a SANS setup at a central wavelength of 3 nm \cite{Flauger-p19}. The maximum first order diffraction efficiencies were lower than $\eta_1<4$\% for a thickness of $d\approx 85\,\mu\text{m}$ and we can give an upper estimate for $\bcr_1<2\,\mu\text{m}^{-2}$ which is definitely not worth to proceed further.
\section{MATERIALS: OTHER}
During the last decade we also investigated other methods than optical holography that would allow a preparation of periodic structures as neutron optical elements. One major advantage of holographic recording is that transport of matter occurs massively and in the \textbf{volume} of the material. This is an inevitable asset for microscopic thicknesses (or heights) at submicron grating spacings which are required for our aims. The high aspect ratios ($\geq 40$) were not achieved yet by other methods, e.g., two-photon lithography of photoresins which is commercially available. 

However, one limitation of the holographic method becomes apparent whenever the material is absorbing light or even opaque. In this case the interference pattern penetrates into the material only over a short depth. This was the case for high nanoparticle concentrations of say nanodiamonds. The same is true for an NPC with superparamagnetic nanoparticles (maghemite) \cite{Gyergyek-csa08,Klepp-jpcs12}. They are an interesting species as neutrons are sensitive to magnetic moments because of their spin (magnetic scattering). If gratings with superparamagnetic nanoparticles could be fabricated, their diffractive properties could be altered by application of an external magnetic field ("magnetically switchable gratings"). 

Therefore, a first attempt to overcome this problem was made by first preparing standard holographic polymer dispersed LC gratings where the LC component rich regions should serve as interchangeable fillers; next the LCs were to be removed leaving back an empty (periodic) scaffold of polymer; finally these voids would be refilled with the desired nanoparticle suspension. Experimental tries did not succeed.

A second attempt was made by a completely different technique: the self-assembly of colloidal polystyrene particles with superparamagnetic nanoparticles in the suspension to from a colloidal crystal. Its thickness was around then micrometers, just enough to observe noticeable neutron diffraction of less than $\eta<1$\% in a SANS setup \cite{Licen-jpcs17}. Investigating the diffraction of neutrons with and without application of a magnetic field unfortunately showed only a small influence if any.

After all, a novel, sophisticated nanotechnological technique deems extremely promising. By employing a sequence of production steps combining optical lithography with PVD, metal assisted chemical etching finally followed by a smart drying step the authors succeeded to fabricate gratings, coined by them "nanogratings", with a high aspect ratio ($>40$) \cite{Michalska-afm23}. Their applicability for x-rays was demonstrated. This might be a route to follow also for VCN.

\section{Summary and Conclusion}
The progress of fabricating efficient and versatile gratings as neutron optical elements suitable for VCN interferometry was briefly reviewed. A particular focus was laid on the materials aspects. Optical holographic methods are suitable for making such artificial structures in polymers and polymer composites with highest coherent scattering length modulations of about $\bcr\approx10\,\mu\text{m}^{-2}$. At the required moderate thicknesses (i.e. low angular selectivity) of a few tens of microns diffraction efficiencies of 30\% - 70\% can be reached at very cold neutron wavelengths. The design and setup of a VCN interferometer based on such gratings is under way.

Other novel nanotechnological methods that have proven to act as very efficient diffractive elements for X-rays might also be appropriate for VCN. Tests on these structures are ongoing.

\acknowledgments 
 
The hospitality of the ILL and technical assistance by Thomas Brenner at the very cold neutron beamline PF2/VCN is acknowledged. E.H. is grateful for financial support by the Vienna Doctoral
School in Physics (VDSP), in particular, a VDSP mobility fellowship. We thank Dr. Bruder and Covestro for providing us with the \bay material, Mostafa A. Ellabban for preparing the ionic liquid - polymer composites and Sonja Falmbigl for making available unpublished first light optical data on the second harmonics of the grating in CAS shown in Fig.\,\ref{fig:DE2}.

This research was funded in part by the Austrian Science Fund (FWF) [P-35597-N], the Austrian Research Promotion Agency (FFG), Quantum-Austria NextPi, grant number FO999896034 and the European Union: "NextGenerationEU".


\end{document}